\documentstyle[psfig,conf_iap,]{article}
\begin{document}

\heading{Dark energy effects in the Cosmic Microwave Background Radiation} 
\par\medskip\noindent
\author{Pier Stefano Corasaniti$^{1}$}
\address{Centre for Theoretical Physics, University of Sussex, 
BN1 9QH, Falmer, Brighton, United Kingdom}

\begin{abstract}
Understanding the nature of the dark energy is one of the 
most important task in cosmology. In principle several 
cosmological observations can be used to discriminate amid 
a static cosmological constant contribution and a dynamical
quintessence component. In view of the upcoming high
resolution CMB experiments, using a model independent
approach we study the dark energy imprint in the CMB 
anisotropy power spectrum and the degeneracy with other
cosmological parameters.
\end{abstract}

\section{Dark energy models}
Two competitive scenarios have been proposed
to take into account for the observed accelerated expansion of the
universe \cite{PERL}, the cosmological constant and a plethora of
`quintessence' scalar field models.
It has been recently shown that an appropriate parameterization of the time 
behavior of the dark energy equation of state, can reproduce the dynamics of
several quintessence potentials and many other models for the acceleration
\cite{CORAS}. In particular different models
are specified by the value of the equation of state today $w_Q^o$, its
value during the matter dominated era $w_Q^m$, the value
of the scale factor $a_c^m$ where the equation of state changes from
$w_Q^o$ to $w_Q^m$ and the width $\Delta$ of this transition. 
We refer to the formula (4) of the equation of state in reference \cite{CORAS}.
We may distinguish two different classes of dark energy models,
those with a rapid transition, characterized
by the ratio $a_c^m/\Delta>1$,
and those with a slowly varying behavior, for which $0<a_c^m/\Delta<1$.
For instance the `Albrecht-Skordis' model \cite{ALB} belongs to the former
class while the latter includes the inverse power law potential \cite{ZAT}.
We shall study the dark energy effects in the CMB power spectrum and
we will discuss the possibility to detect these imprints with the next generation of CMB
experiments.

\section{Acoustic peaks}
The angular size of the sound horizon at the decoupling sets the location of
the acoustic peaks. Different dark energy models can lead to a different
angular diameter distance to the last scattering surface and consequently
to a shift of the peaks.
For instance if $w_Q^o$ is close to the cosmological constant value,
the universe starts to accelerate at earlier times, the distance to the last scattering
surface is farther and the peaks are shifted toward larger multipoles. 
The opposite effect occurs if $w_Q^m$ tends to the dust value or if $a_c^m$ tends to
its present value $a_o=1$, but in these cases the amplitude of the shift
depends on the value of $\Delta$.
In fig.1 we plot the relative difference to a $\Lambda$CDM model of the position
of the first, second and third peaks for rapidly varying models (fig.1a-1b) and
slowly varying ones (fig.1c-1d). We set $w_Q^o=-1$.
As we can see, in models with a rapidly varying equation of state
the discrepancy with respect to the $\Lambda$CDM increases for $a_c^m \approx 1$ (fig.1a),
 while for slowly varying behaviors of the equation of state there is not
dependence on $a_c^m$ (fig.1c).
 This is what we
expect since in the latter class of models a different value of $a_c^m$
does not affect the evolution of the dark energy density. 
We notice that in general the shift of the position
of the first peak is always larger than that of the second and third ones.
This is due to the ISW effect that is a characteristic signature of dark energy
at low multipoles.
\begin{figure}
\centerline{\vbox{
\psfig{figure=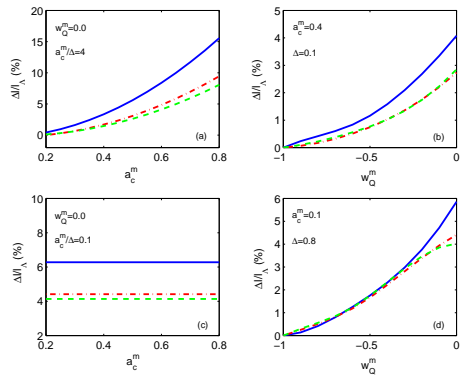,height=5.cm}
}}
\caption[]{Relative difference of $l_1$ (blue solid line), $l_2$ (green dashed
line) and $l_3$ (red dash-dotted line) to the $\Lambda$CDM model, for rapidly
varying models ($top$ $panels$), and with slow transition ($bottom$ $panels$).
For these models the present value of the equation of state is $w_Q^o=-1$.
}
\end{figure} 
The location of the Doppler peaks is an efficient tool to test the dark energy
\cite{DORAN}, however we should consider the degeneracy with other cosmological
parameters. For example, for small values of $H_o$ the peaks are shifted toward large
multipoles. The same effect occurs for large values of the physical baryon density
$\Omega_b$, since the size of the sound horizon becomes smaller. Therefore we 
expect the value of the dark energy parameters
to be degenerate mainly with $H_o$ and $\Omega_b$.
\section{ISW effect}
The ISW produces late time anisotropies at the large angular scales
of the CMB and is caused by the decay of the gravitational potentials \cite{REES}.
In a $\Lambda$CDM model such a decay starts as soon as the universe enters
in the cosmological constant dominated phase. On the other hand
the quintessence contributes to this decay not only with the dynamics of
the background but with its clustering properties as well. 
In fact the quintessence is not a smooth component,
it has fluctuations that can cluster at the very large scales 
and modify the evolution of the matter perturbations. Therefore
the imprint of the dark energy in the ISW depends
on its specific features \cite{CORA}. In fig.2 we plot the power spectrum $C_l^{ISW}$ for 
the models of fig.1, the red line corresponds to the ISW contribution in a $\Lambda$CDM
model. As we can see, for rapidly varying models, there is a boost of power
for transitions occurring at close redshifts (fig.2a). The $C_l^{ISW}$
then decreases toward the cosmological constant case as the transition moves at higher redshifts (fig.2a).
The largest amplitude is obtained for $w_Q^m=0.0$, but the signal is suppressed for
$w_Q^m<0.0$ (fig.2b). In the case of slowly varying models the ISW
is almost independent of $a_c^m$ (fig.2c). It is larger than the $\Lambda$CDM
for $w_Q^m=0.0$, becomes smaller for negative values of $w_Q^m$ and tends
to the $\Lambda$ case for $w_Q^m<-0.3$ (fig.2d). 
The result of this qualitative analysis suggests that amid the dark energy models with
$w_Q^o$ close to $-1$, those with a rapid transition occurring at low redshifts
produce the most distinctive signature in the CMB power spectrum.
On the contrary those with a slowly varying behavior are more difficult to 
distinguish from a cosmological constant model. 
\begin{figure}
\centerline{\vbox{
\psfig{figure=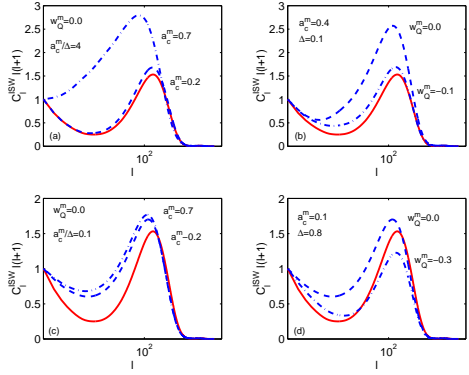,height=5.cm}
}}
\caption[]{Power spectrum of the ISW for rapidly varying models ($top$ $panels$) and
slowly varying ones ($bottom$ $panels$). The solid red line is the ISW effect produced
in $\Lambda$CDM case.
}
\end{figure} 

\section{Ideal CMB measurements}
In order to test the sensitivity of an ideal CMB experiment to the dark energy
effects, we simulate a sample of $C_l$s with cosmic variance errors.
We assume a fiducial model with a slowly varying equation of state with
the following parameters: $\Omega_{Q}=0.68$,
$H_o=70$ Km $s^{-1}$ $Mpc^{-1}$, $w_Q^o=-1$, $w_Q^m=-0.4$, $a_c^m=0.2$, $\Delta=0.3$. 
We assume a flat geometry, no tensor
contribution, the scalar spectral index $n=1$ and the baryon density $\Omega_b h^2=0.021$.
We assume a gaussian prior on $\Omega_m=1-\Omega_Q$ with $\sigma_m=0.05$.
We perform a likelihood analysis over $\Omega_Q$,
$H_o$, $w_Q^o$, $w_Q^m$ and $a_c^m$.
The results are shown in fig.3. As we may note, unless we take a prior on $H_o$,
the value of $w_Q^o$ is poorly constrained. More important is the likelihood plot
in the $w_Q^m-a_c^m$ plane. As we expect,
$a_c^m$ is undetermined. On the other hand $w_Q^m$ is not
very well constrained even with the $H_o$ prior. This is because $w_Q^o$ and $w_Q^m$
are degenerate, hence marginalizing the likelihood over $w_Q^o$ shifts
the best fit value of $w_Q^m$.
In particular we cannot exclude the case $w_Q^m=-1$. Therefore
we conclude that if the present value of the dark energy equation of
state is close to $-1$, a large class of models will not be distinguished
from a $\Lambda$CDM scenario even with ideal CMB measurements. 
It is possible that combining different cosmological data, as SN Ia, large
scale structure and quasar clustering we can break the degeneracy and
infer more information on this class
of dark energy models.
\begin{figure}
\centerline{\vbox{
\psfig{figure=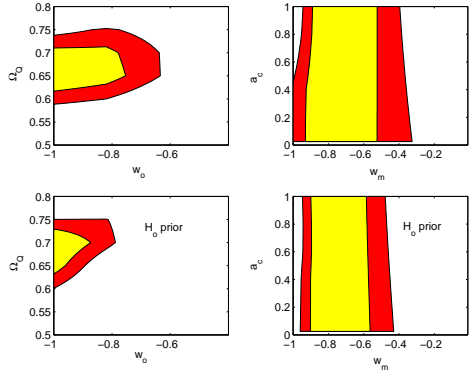,height=5.cm}
}}
\caption[]{Simulated 2-D likelihood contour plots for an ideal experiment,
with no priors ($top$ $panels$) and $H_o$ prior ($bottom$ $panels$). 
The yellow and red contours correspond to $1$ and $2\sigma$ respectively.
}
\end{figure}

\acknowledgements{We are grateful to Bruce Bassett, Carlo Ungarelli and
Ed Copeland,
with whom much of the work reviewed here has been done. PSC is suported by a University of Sussex bursary.}

\vfill

\begin{iapbib}{99}{
\bibitem{PERL} Perlmutter, S et al., 1999, \apj 517, 565
\bibitem{CORAS} Corasaniti, P. S. and Copeland, E. J., 2002, astro-ph/0205544
\bibitem{ALB} Albrecht, A., and Skordis, C. 1999, Phys. Rev. Lett. 84, 2076
\bibitem{ZAT} Zlatev, I., Wang, L., and Steinhardt, P. J. 1999, Phys. Rev. Lett. 82, 896
\bibitem{DORAN} Doran, M., Lilley, M. J., 2002, Mon. Not. Roy. Astron. Soc., 330, 965
\bibitem{REES} Rees, M. J. and Sciama, D.W. 1968, Nature 217, 511
\bibitem{CORA} Corasaniti, P. S., Bassett, B. A., Ungarelli, C. and Copeland, E. J., astro-ph/0210209
}
\end{iapbib}
\end{document}